# All-voltage control of Giant Magnetoresistance

Lujun Wei[1,2,*], Yiyang Zhang[2], Fei Huang[1], Jiajv Yang[1], Jincheng Peng[1], Yanghui Li[1], Yu Lu[2], Jiarui Chen[2], Tianyu Liu[2], Yong Pu[1,*], Jun Du[2,*]

[1]School of Science, Nanjing University of Posts and Telecommunications, Nanjing 210023, China.

[2]National Laboratory of Solid State Microstructures and Department of Physics, Nanjing University, Nanjing 210093, China.

[*] Corresponding Authors: wlj@njupt.edu.cn, yongpu@njupt.edu.cn, jdu@nju.edu.cn.

## Abstract

The aim of voltage control of magnetism is to reduce the power consumption of spintronic devices. For a spin valve, the magnetization directions of two ferromagnetic layers determine the giant magnetoresistance magnitude. However, achieving all-voltage manipulation of the magnetization directions between parallel and antiparallel states is a significant challenge. Here, we demonstrate that by utilizing two exchange-biased Co/IrMn bilayers with opposite pinning directions and with ferromagnetic coupling through the Ruderman-Kittel-Kasuya-Yosida interaction between two Co layers, the magnetization directions of the two ferromagnetic layers of a spin valve can be switched between parallel and antiparallel states through all-voltage-induced strain control. The all-voltage controlled giant magnetoresistance is repeatable and nonvolatile. The rotation of magnetizations in the two Co layers under voltages, from antiparallel to parallel states, occurs in opposite directions as revealed through simulations utilizing the Landau-Lifshitz-Gilbert equation. This work can provide valuable reference for the development of low-power all-voltage-controlled spintronic devices.

**Introduction**

Efficient 180° magnetic switching through electrical means is critical for spin-based data storage and logic[1,2]. However, current methods mainly rely on local magnetic fields or spin torques[3-5]. Voltage-controlled magnetization switching is preferred due to its lower energy consumption[6], but it is challenging to achieve because voltages cannot induce the necessary time-reversal symmetry breaking for 180° magnetization switching. Various approaches, such as magnetoelectric coupling in magnetoelectric materials [7-10] and a solid-state ions implantation in ferrimagnet materials[11-14], are being explored for voltage-controlled magnetization switching, but they often require either complex fabrication procedures or high electric fields (more than 100 kV/cm) to induce sufficient ions.

Extensive research has focused on utilizing voltage-induced strain to switch the magnetization[15-25]. The magnetization direction in a ferromagnetic (FM) material is often sensitive to applied strain, which can make FM magnetization rotation. This phenomenon, known as magnetoelastic or inverse magnetostriction effect, has been widely investigated for its strain-induced change in magnetic anisotropy in FM thin films. When voltage-controlled magnetization switching is used in practical scenarios, it is typically integrated into magnetic devices. For a traditional spin valve, the resistance of a magnetic trilayer structure, consisting of two FM layers separated by a non-magnetic (NM) metal layer, depends on the relative angle between the magnetizations of the two FM layers. One layer is the free layer, the other is the reference layer pinned by the antiferromagnetic layer adjacent to it. Applying an external magnetic field modifies the relative angle, resulting in a resistance change that is useful for sensing magnetic field magnitude or direction. Through this voltage-controlled strain-induced mechanism, a FM layer can be only switched its magnetization by 90°, meaning the angle of magnetizations of the two FM layers varies from 0° to 90°[15,17,26-30]. A very complex twice strains needs to be applied if a 180° magnetization switching is obtained without external field assistance[18]. Hence, magnetizations switching of a device between parallel and antiparallel states using a voltage control is a highly challenging task.

In this work, we introduced an approach to achieve all-voltage-controlled magnetization switching between parallel and antiparallel states of a spin valve. We departed from traditional device design schemes by employing two FM layers, both of which are free layers sensitive to voltage-induced strain. By satisfying two conditions: (1) arranging the two FM layers antiparallel, which can be achieved through antiferromagnetic and FM exchange coupling. Because cobalt film has a large negative magnetostrictive coefficient and is sensitive to compressive strain, it is chosen as the FM layer. (2) existing indirect FM coupling between the FM layers, which can be accomplished by adjusting the thickness of the NM (or space) layer (e.g., Cu) between them, specifically tuning the Ruderman–Kittel–Kasuya–Yosida (RKKY) interaction[31,32], as shown in Fig. 1a. Such spin valve corresponding to the initial state (IS) is a high resistance state (HRS). The magneto-elastic properties of the Co layer grow on ferroelectric substrate is manipulated by voltage-induced in-plane compressive strain, making its strain direction more preferable for the magnetization arrangement. This can be achieved by using two FM layers rotated by 90° to the parallel direction under a voltage-induced tensile strain (Fig. 1b), as the two Co layers were ferromagnetically coupled to each other, resulting in a low resistance state (LRS). The device returns to the IS at the voltage.

We have successfully achieved all-voltage-controlled magnetization switching between parallel and antiparallel states of a spin valve, which provides a reference for precise experimental design of spintronic devices. Through simulations, we found that the two FM layers move from the antiparallel state to the parallel state with opposite progression.

**Results**

    **Structure and magnetism of device.** The stacking structure of device is shown in Fig. 2a, with a Ta/IrMn/Co/Cu/Co/IrMn/Pt/Ta/SiO$_x$/Ta/Cu/Ta multilayer film deposited on a ferroelectric single crystalline (011)-oriented [Pb(Mg$_{1/3}$Nb$_{2/3}$)O$_3$]$_{0.68}$-[PbTiO$_3$]$_{0.32}$ (PMN−PT) substrate using sputtering (see Methods and Supplementary Fig. S1). A Ta/Cu/Ta trilayer serves as the top electrode, while another is the bottom electrode grown on the PMN-PT serve as the back electrode for using to apply voltage-induced strain. The SiO$_x$ layer acts as an insulating layer to achieve uniform strain for applied to the multilayer film and to make the magnetoresistance (MR) test and the voltage applied to PMN-PT independent. During the growth of the bottom Co/IrMn and top IrMn/Co multilayers, opposite magnetic fields were applied and paralleled to the $[01\bar{1}]$ direction of the PMN-PT, respectively, so that the two layers of Co were pinned in opposite directions with opposite exchange biases.

    To investigate the structural properties of device, cross-sectional scanning transmission electron microscopy (STEM) was used. Figure 2b shows a high-angel annular dark-field (HAADF) STEM image of the multilayer film, exhibiting a good stacking structure. The interfaces between Ta and Pt, Pt and IrMn, IrMn and Co, and SiO$_x$ and Ta are very sharp (Fig. 2c), while interfaces between Cu and two Co layers are ambiguous. Because the atomic numbers of Co and Cu are so close, so STEM cannot better distinguish them. To further detect those element, line profiles of the atomic counts were measured via energy-dispersive X-ray (EDX) spectrometry (Fig. 2d). The peaks of Cu and Co can be clearly observed, and the peak of Cu atoms is

located between two Co atoms peaks, indicating it exists a Co/Cu/Co structure. Bright-filed cross-sectional TEM images and the mapping of these elements in the multilayer film (see supplementary Fig. S2) exhibit the stacking structure with the IrMn/Co/Cu/Co/IrMn/Pt/Ta and good crystallinity for the IrMn, Co and Pt layers, providing good conditions for building exchange bias effect between Co and IrMn layers.

Figure 2e show a double normalized magnetic hysteresis (*M-H*) loop of the multilayer film by focus magneto optical Kerr effect (FMOKE) measurements. The normalized *M-H* loops in positive and negative fields are respectively correspond to magnetic properties at the top Co/IrMn and the bottom IrMn/Co of the multilayer film. The exchange bias fields ($H_{EB}$)/coercivities ($H_C$) of the positive and negative *M-H* curves are 67.2/18.5 Oe and -100.0/54.2 Oe, respectively. The exchange bias field ($H_E$) and coercivity ($H_C$) are defined as $H_{EB} = (-H_{CL} - H_{CR})/2$ and $H_C = (H_{CL} - H_{CR})/2$, respectively. Here, $H_{CL}$ and $H_{CR}$ are the coercive fields for the descending and ascending branches of the positive and negative *M−H* curves, respectively. The magnetizations of two Co layers with opposite exchange bias fields are opposite directions at *H* = 0 Oe. Moreover, a *M-H* loop of the control sample Ta/Co/Cu/Co/IrMn/Pt/Ta without the top IrMn layer under the same preparation conductions as the device (the inset of Fig. 2e) show that two Co layers are ferromagnetically coupled to each other due to the fact that the *M-H* curve of the top Co layer is shifted to the negative in the RKKY interaction, and the coupling field is about 19.5 Oe.

**All-voltage control of giant MR.** The measuring configuration is shown in Fig. 3a, and MR curves were measured using magnetic field *H*//*y* axis to control voltages. Prior to measuring MR curves, we conducted a tested on the voltage-induced strain characteristics of the PMN-PT using a commercial strain gauge (the inset of Fig. 3b). The results demonstrated an obvious strain-voltage (*S-V*) hysteresis loop along the [$01\bar{1}$] direction of the PMN-PT (Fig. 3b). Under negative voltages, the PMN-PT exhibited tensile strain in the [$01\bar{1}$] direction, while it exhibited compressive strain in the [100] direction (see supplementary Fig. S3). The PMN-PT displays anisotropic and significant remanent strain, with remanent tensile strain of approximately 1.5% in the [$01\bar{1}$] direction. The voltage-induced strain results from a change of the lattice constant, which is proved by offsetting the (022) diffraction peaks of the PMN-PT using X-ray diffraction (XRD) under *in-situ* voltages (supplementary Fig. S4).

To investigate the effect of voltage-induced strain on magnetic anisotropy, we grew a Ta/Cu/Ta/SiO$_x$/Ta/Pt/Co/Ta multilayer film on the PMN-PT substrate, and conducted an FMOKE test under *in-situ* voltages (see supplementary Figs. S5). The magnetization direction of the Co layer rotates 90 degrees from [$01\bar{1}$] to [100] directions under a negative voltage, and it rotates back to 0 degrees from [100] to [$01\bar{1}$] directions under a positive voltage. Furthermore, various magnetizations of the Co layer under voltages (Supplementary Fig. S6) are consistent with the voltage-induced strain character (see Fig. 3b and supplementary Fig. S3).

MR curves were measured at room temperature under various voltages in the *H//y* axis (Figs. 3c-f and supplementary Fig. S7) to investigate the voltage effects on the spin valve. Initially, a fixed forward voltage of 300 V (6 kV/cm) was applied to the PMN-PT substrate and the MR curve was obtained (Fig. 3c), revealing a butterfly-type curve with an MR of approximately 3.5%. Here, the MR ratio is defined as $\frac{R(0)-R(H)}{R(H)} \times 100\%$. Subsequently, as the voltage decreases from +300 to 0 V (denoted as +0 V), the MR value reaches a maximum of approximately 3.6% (Fig. 3d). The MR of the device at +300 V is slight smaller than that at +0 V due to the higher tensile strain (0.03%) in the $[01\bar{1}]$ direction at 300 V compared to that (0%) at +0 V. When the applied *H* is between -50 and 50 Oe at +0 V, the Co magnetization of the two layers in the device is basically in an anti-parallel state, and when the field is increased to 150 Oe or decreased to -150 Oe, it is in a parallel state. Subsequently, when a negative voltage of -300 V was applied, the voltage-induced the MR curves has a huge change, i.e. the butterfly-type MR curve disappear and it is almost close to overlap together, resulting in a minimum MR of approximately 0.8% (Fig. 3e).

When a spin valve is rotated in a constant applied field, the resistance as a function of *θ*, the angle between the magnetizations of the two FM layer, is described by the following relationship [33]:

$$R(\theta) = 0.5(R_{AP} - R_P) - R_P + 0.5(R_{AP} - R_P)\cos(\theta), \qquad (1)$$

where $R_P$/$R_{AP}$ represent the resistance when the magnetizations of two Co layers are parallel/anti-parallel to each other. It was estimated that the corresponding angle at zero magnetic field is approximately 173 degrees, indicating that the magnetizations of the two Co layers are almost anti-parallel, demonstrating the transition of the two Co layers of magnetizations from a parallel state to an anti-parallel state under -300 V. Finally, the MR value is 1.4% at -0 V (Fig. 3f), and it was estimated that the corresponding angle between the Co layers at *H* = 0 Oe is about 120 degrees due to remain strain.

The MR ratios versus the fixed varies voltages (Fig. 3g) demonstrate a hysteresis loop that is oppositive with the varying voltage-induced strain in the $[01\bar{1}]$ direction (Fig. 3b), and is consistent with that in the [100] direction. Furthermore, the MR curves of the device were measured by applying both the magnetic field and the current along the *x* direction (see supplementary S8). The relationship between MR and voltages also exhibits hysteresis loops (supplementary S9).

The MR ratio versus cycle number was obtained as the device was cycled more than 100 times between the -0 V, +0 V, 300 V and -300 V states (Supplementary Fig. S10a). The results indicate that the MR switching is remarkably reversible, with no evidence of device wear-out or irreversible changes of the magnetic property after an unprecedented degree of cycling. More importantly, a distinct hysteresis loop of voltages control of device resistance (Fig. 3h) show the different resistances at $H = 0$ Oe for +0, -0, 300 and -300 V states. These resistances were stable (see supplementary Fig. S10b), which presents the resistance change of the device at +0, -0, 300 and -300 V, respectively, and demonstrates nonvolatile voltage-controlled resistance. Applying a voltage of +0 V induced an HRS, whereas applying a voltage of - 0 V induced an LRS.

The resistance-voltage (*R-V*) curves were measured in a fixed magnetic field along the *y* axis (Supplementary Fig. S11) and *x* axis (Supplementary Fig. S12), respectively. The curves with anti-clockwise direction remained essentially unchanged within the range of -50 to 50 Oe with *H*//*y* axis (or 100 Oe with *H*//*x* axis), indicating the device's high stability. The size of the *R-V* loops decreased as the fixed magnetic field decreased from -50 to -70 Oe and increased from 50 to 100 Oe for *H*//*y* axis, respectively (Supplementary Fig. S11). When the magnetic field minimized -80 Oe or exceeded 125 Oe, the curve transitioned into a clockwise loop. This change occurred because the applied magnetic field is greater than the exchange bias field, aligning the magnetization of the two Co layers in a parallel along the *x* axis for the IS state. When the applied voltage is -300 V, the magnetizations of them are rotation together. As the magnetic field continued to decrease in negative or increase in positive, the loop size further decreased due to the large external magnetic field overpowering the voltage-induced anisotropy field effect, preventing the rotation of magnetizations. The loop formation is attributed to the multilayer itself of stretching and compression. Similar characteristics occur in the $H//x$ axis (Supplementary Fig. S12).

**Simulations of dynamic rotation processes for two ferromagnetic layers.** The behavior of voltage-induced the magnetic moments progression of two Co layers was studied using the Landau-Lifshitz-Gilbert (LLG) equation as follows [34].

$$\frac{d\boldsymbol{m}}{dt} = \gamma \boldsymbol{m} \times H_{eff} + \alpha \boldsymbol{m} \times \frac{d\boldsymbol{m}}{dt}, \qquad (2)$$

Where $\gamma$ is the gyromagnetic ratio, $\alpha$ is the effective damping parameter and a value of $\alpha = 0.01$ [35]. In an assuming single domain system, the macrospin of Co is represented by a cosine vector of uniform magnetization $\boldsymbol{m} = \boldsymbol{M}/M_S = (m_x, m_y, m_z)$, and the magnetization state in the cell is affected by the effective anisotropy field ($H_{eff}$). The $H_{eff}$ is expressed as follows [22,36],

$$H_{eff} = -\nabla_m \left( \frac{1}{2} H_z m_z^2 + \frac{1}{2} H_\sigma^V m_x^2 + \frac{1}{2} H_C m_y^2 - (\boldsymbol{H_R} + \boldsymbol{H_{EB}}) \cdot \boldsymbol{m} \right), \qquad (3)$$

Here, the out-of-plane effective anisotropy field ($H_z$) is approximately 17 kOe as demonstrated by previous reports [37]. The value of $H_C$ of the top and bottom Co layers is approximately 18.5 and 54.2 Oe, respectively. The RKKY field ($H_R$) is approximately 19.5 Oe in the magnetization parallel of the Co layers. The value of $H_{EB}$ of the top and bottom Co layers is approximately 67.2 and -100 Oe, respectively, in $y$ axis direction. The voltage effect is introduced as a change in the anisotropy field ($H_\sigma^V$), which is assumed to be approximately 460 Oe along the $x$ axis (or [100] direction) under a voltage of -300 V (for details, see Supplementary Note 1). The value of the $H_\sigma^V$ is 0 Oe at +0 V because the corresponding strain is zero.

The relaxation process of $M/M_S$ with a voltage (Fig. 4a) indicates that if the width of the voltage pulse ($t$) is more than 3 ns, magnetization rotation can be achieved, with a response time of less than 10 ns, which is comparable to the writing time of the typical random access memories [38]. The magnetization $m_1$ of the top Co layer and $m_2$ of the bottom Co layer are respectively oriented along the $+y$ and $-y$ directions, denoted as IS or anti-parallel state (Figs. 4b and c). Then a voltage of -300 V is applied, resulting in the rotation of $m_1$ ($m_2$) from $+y$ ($-y$) to $-x$ directions due to the increase of the in-plane effective anisotropy field in the $x$ direction. As a result, the magnetization rotates to $-x$ axis nearly, denoted as final state (FS) or parallel state (Figs. 4b and c). When the voltage is switched off and sets to +0 V, the magnetization returns to IS after the relaxation process (Figs. 4d and e). Note that the trajectories of the rotation processes between the IS to the FS are opposite for the $m_1$ and $m_2$.

**Conclusion**

In conclusion, we demonstrated a reversible, coherent rotation of magnetization and manipulation of Giant MR at room temperature in the IrMn/Co/Cu/Co/IrMn/PMN-PT structure using voltages, without the need for any external magnetic field. These effects are attributed to the change in magnetic anisotropy resulting from the strain anisotropy of PMN-PT under external voltages. The simulated trajectories of magnetization rotation, tuned by voltages, show that the two Co layers switch between anti-parallel and parallel states, resulting in opposite direction rotation. We believe that our results have significant technological implications, and our approach provides a new method for magnetization switching in all-voltage-controlled spintronics devices. Finally, if a CoFeB/MgO/CoFeB tunnel junction is fabricated using this method, and the symmetry of voltage-induced strain-regulated magnetization rotation is broken by the dipole-dipole interaction between the two FM layers, the voltage control of MR value will experience a huge increase, rendering the device more suitable for practical applications.

**Acknowledgements**

This work was supported by the National Natural Science Foundation of China (Grant Nos. 52001169, 51971109, 61874060, U1932159, 61911530220, and 12104238), the open Project of the laboratory of Solid-state Microstructures of Nanjing University (Grant No. M36036).

**Methods**

**Sample preparation.** The GMR devices were deposited on (011)-oriented PMN-PT substrates by magnetron sputtering at room temperature. The top (Ta(5)/Cu(50)/Ta(5)) and bottom (Ta(5)/Cu(100)/Ta(5)) electrodes were respectively deposited on a positive polishing surface and a unpolished surface on back of the PMN-PT substrates, respectively. A SiO$_x$(100 nm) film were then grown on the top electrode. The sample is excited over 10 times by applying a voltage, which causes it to stretch and compress, and finally set to a compressed state in the $[01\bar{1}]$ direction of a PMN-PT substrate. Following, the Ta(3)/IrMn(8)/Co(8)/Cu(1.85)/Co(6.4)/IrMn(8)/Pt(5)/Ta(2) multilayers formed into a 100 μm-wide and 2 mm-long stripe using the metal mask, and Ta/Co(8)/Cu(1.85)/Co(6.4)/IrMn/Pt/Ta and Ta(3)/Co(6.4)/IrMn(8)/Pt(5)/Ta(2) multilayers formed into a 100 μm-diameter disk were deposited on the SiO$_x$ film upper surface. The units of the numbers in parentheses are nanometers. A disk shape was chosen to avoid influences from the shape anisotropy of a Co layer. During the growth of Co/IrMn and IrMn/Co bilayers, magnetic fields of approximately 100 Oe were applied in opposite directions, respectively.

**Experimental set-up.** The sample were with probes for four-terminal measurements. The MR curves were measured under a direct current of 0.5 mA. In-plane external magnetic fields were applied using an electromagnet. The in-plane magnetic hysteresis (*M–H*) loops were measured by a home-made focus magnetic optical Kerr effect (FMOKE) magnetometer with a 532 nm green laser light with a spot size of about 5 µm. The crystal structure was studied by a Bruker D8-Advance XRD using Cu Kα radiation (λ =0.154 nm) under *in-situ* voltages. The TEM, STEM and EDX measurements were conducted on an FEI Themis Z aberration-corrected TEM equipped with a Super ×4 silicon drift detector, operating under a 300 kV accelerating voltage. All measurements and characterizations were conducted at room temperature.

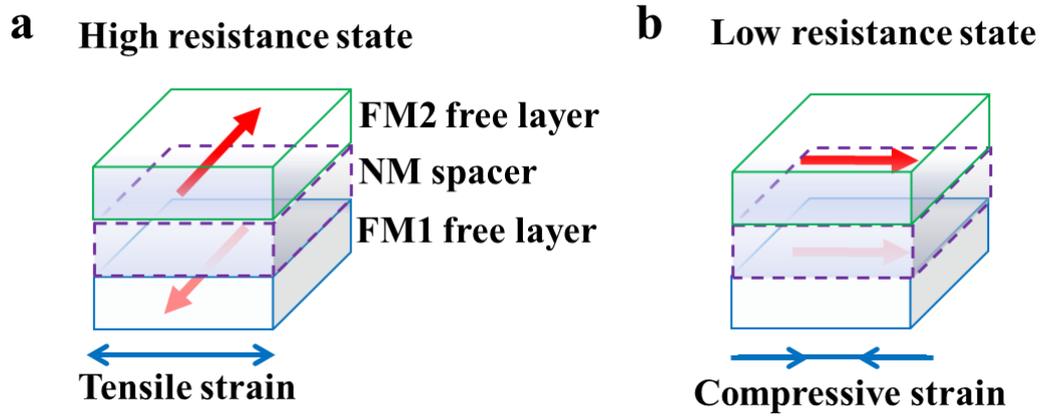

**Fig. 1 (a) Schematic of magnetization state by controlling voltage-induced-strain.** (a) Anti-parallel and (b) parallel arrangements of the magnetization directions for two FM layers under the tensile and compressive strains, respectively.

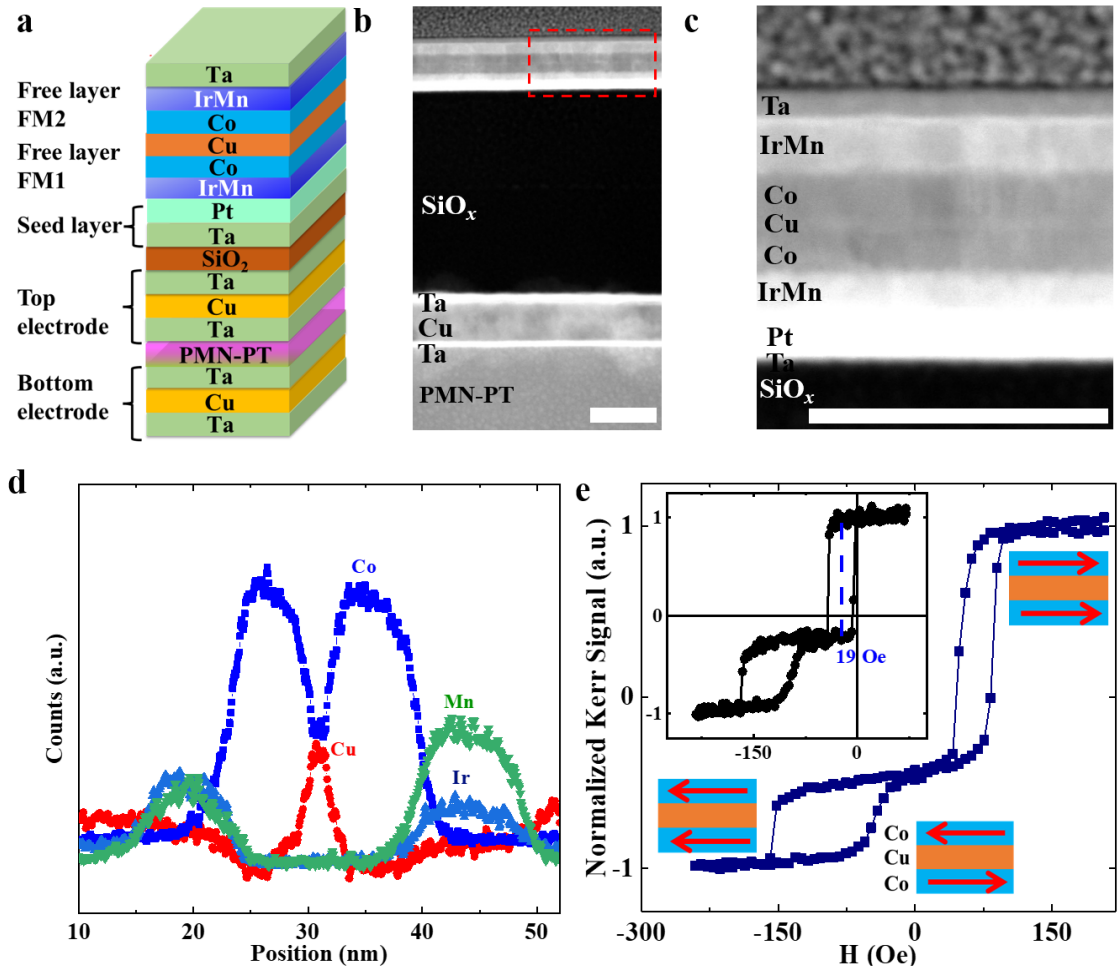

**Fig. 2. Schematic and Magnetism of device.** Schematic (a) and HAADF-TEM (b, c) image of sample's stacking structure. The (c) image corresponding to the locations denoted by the red box in (b). (d) The EDX line profile image of the Ta, Pt, Co, Ir and Mn elements of the top multilayer film. (d) Normalized *M-H* loop of the multilayer film in the $[01\bar{1}]$ direction. The inset is normalized *M-H* loop of a Ta/Pt/Co/Cu/Co/IrMn/Pt/Ta multilayer film. Scale bar is 50 nm.

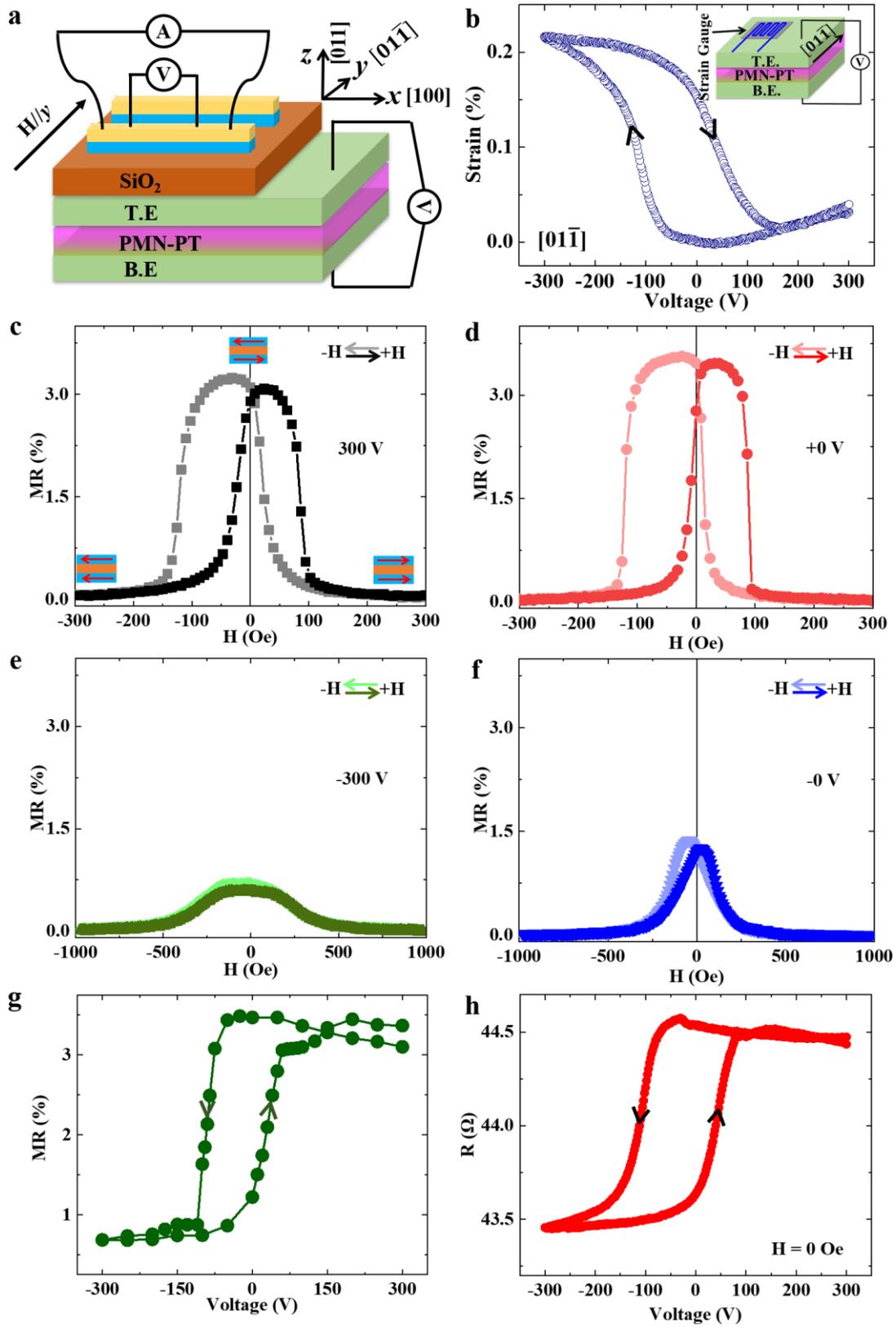

**Fig. 3. Voltage control of magnetoresistance and resistance.** Schemic of measurement device. The applied voltage on the PMN-PT and the measured MR

curves are independent. (b) Strain-voltage curve in the $[01\bar{1}]$ direction. The inset of (b) is the measuring configuration of MR curves. The MR curves of device by applied a voltage of 300 V (c), +0 V (d), -300 V (e) and -0 V (f), respectively, on the PMN-PT substrate. (g) The MR values versus applied fixed varying voltages on the PMN-PT substrate. (h) Dependence of resistance on voltage under $H = 0$ Oe for device.

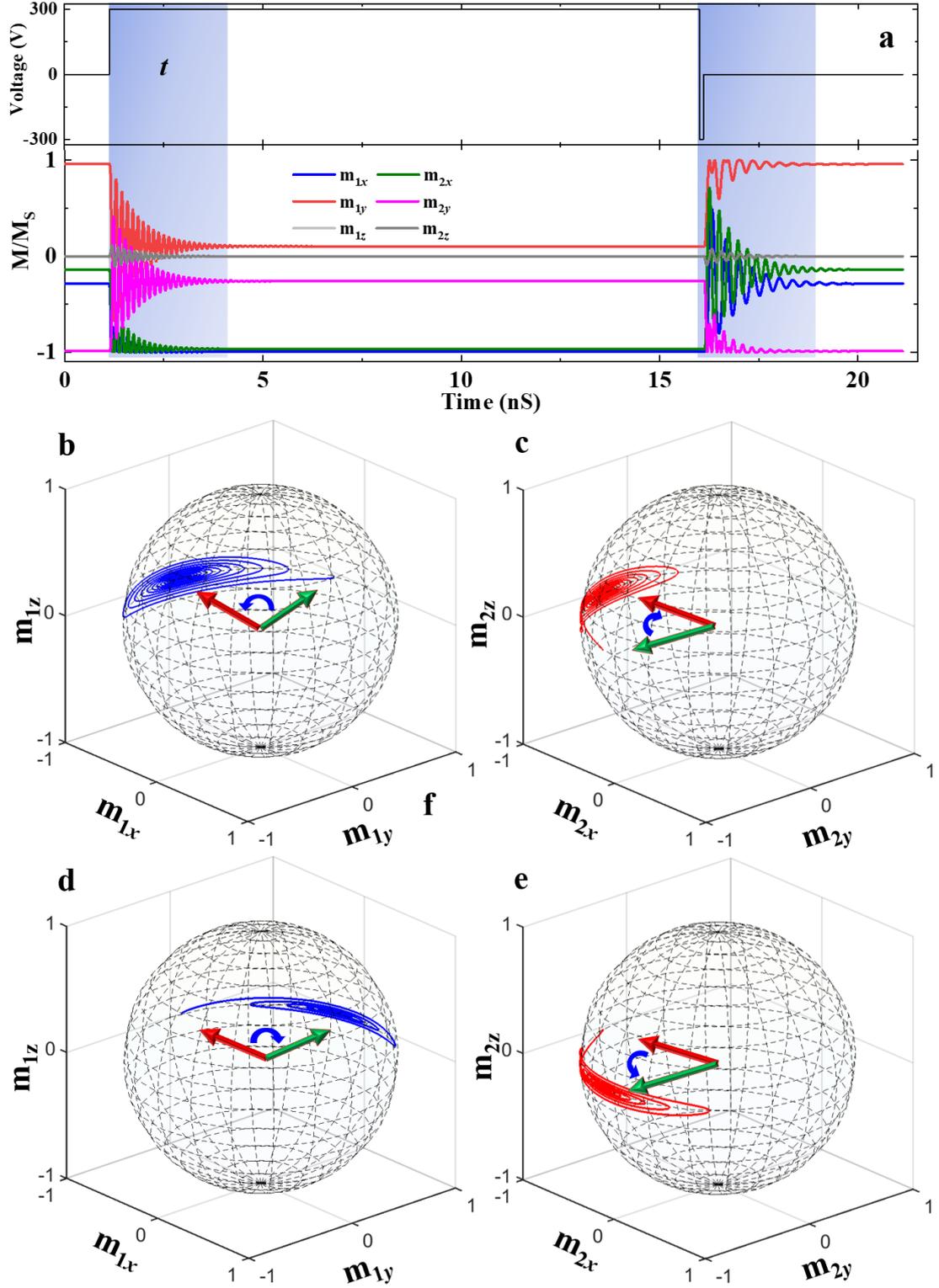

**Fig. 4. Simulation result of the trajectories for processional switching.** (a) Response of $M/M_S$ to a step voltage of -300 V and +0 V. The "$t$" denotes the respond time required for the operating process. The magnetizations in +$y$ direction for $m_1$ (b) and -$y$ direction for $m_2$ (c) are both switched to the -$x$ direction in an in-plane filed $H_\sigma^V$.

The magnetizations of $m_1$ and $m_2$ of in the -$x$ direction are re-backed the +$y$ (d) and -$y$ (e) directions, respectively, after +0 V. The trajectories of the rotation processes are indicated by the blue and red curves in 4b-e. The arrows show magnetizations direction of the IS and FS.